\documentclass[aps,prl,twocolumn, showpacs]{revtex4}
\usepackage{amsmath, amsthm, amssymb,amsbsy,graphicx,times,subfigure,marvosym,color,bm,hyperref}
\usepackage[latin1]{inputenc}
\usepackage{sidecap}

\usepackage{ulem}

\begin{document}

\title{Bose-Einstein Condensation and Many-Body Localization of Rotational Excitations of Polar Molecules}
\author{M. P. Kwasigroch}
\author{N. R. Cooper}
\affiliation{T.C.M. Group, Cavendish Laboratory, University of Cambridge, J. J. Thomson Avenue, Cambridge CB3 0HE, U.K.}


\begin{abstract}

  We study theoretically the collective dynamics of rotational
  excitations of polar molecules loaded into an optical lattice in
  two dimensions. These excitations behave as hard-core bosons
  with a {\it relativistic} energy dispersion arising from the dipolar
  coupling between molecules. This has interesting consequences for
  the collective many-body phases.  The rotational
  excitations can form a Bose-Einstein condensate at non-zero
  temperature, manifesting itself as a divergent $T_2$ coherence
  time of the rotational transition even in the presence of inhomogeneous
  broadening. The dynamical evolution of a dense gas of
  rotational excitations shows regimes of non-ergodicity,
  characteristic of many-body localization and localization protected
  quantum order.

\end{abstract}



\maketitle


The ability to create and control gases of cold polar molecules has
sparked great interest in the quantum many-body physics
associated with long-range and anisotropic dipolar
interactions~\cite{Deiglmayr2008, Ni2008,
  OspelkausPRL2010,OspelkausSC2010, Ni2010,Aikawa2010, Miranda2011,
  Chotia2012}. These systems open up possibilities to create
and to probe interesting many-body phases involving the positional
and/or rotational degrees of freedom of the
molecules~\cite{Baranov2012}. For polar molecules loaded into deep
optical lattices --
with positional motion frozen out -- the rotational excitations can be
used to emulate interesting forms of quantum magnet~\cite{Micheli2006,Lukin2006, Buchler2007, Wall2009, Watanabe2009, 
Yu2009, Wall2010,Trefzger2010,Krems2010,
  Daley2010,Kestner2011,GorshkovPRL,GorshkovPRA}. Recent
experiments~\cite{ColoradoExp} have shown
evidence of the dipole-dipole interactions between molecules in
different lattice sites, which appear as an additional source of
decoherence of rotational
excitations~\cite{ColoradoExp,ColoradoTheory}.

In this paper, we study the many-body physics of the rotational
excitations of polar molecules in a two-dimensional (2D) system.  We
show that in regimes of weak disorder, rather than causing
decoherence~\cite{ColoradoExp,ColoradoTheory}, the dipole-dipole
interactions can in fact  stabilize the coherence up to
arbitrarily long times. This stability of coherence arises from the
formation of a collective many-body phase with true long-range
order.
The essential physics arises from the power-law ($1/r^3$) form of
dipolar interactions between molecules. This coupling causes the
rotational excitations to behave as a gas of (hardcore) bosons with a
{\it relativistic} dispersion $\delta\epsilon_{\mathbf{k}}\propto
|\mathbf{k}|$ at small wavevector $\mathbf{k}$. In contrast to
massive particles ($\delta\epsilon_{\mathbf{k}}\propto |\mathbf{k}|^2$)
this relativistic dispersion allows  
a Bose-Einstein Condensate (BEC) to exist in 2D at non-zero
temperature. We show that the formation of this BEC phase leads to a resistance
to decoherence of the rotational excitations formed by a microwave
pulse, 
even in the presence of inhomogeneities that would broaden the rotational transition for
uncoupled molecules. For very strong inhomogeneous broadening
there is a phase transition into an uncondensed phase for which the
initial coherence decays exponentially in time.

An important feature of the rotational excitations of polar molecules
is that they are effectively isolated from any external heat
bath.  In
the presence of disorder such systems are natural candidates to show
many-body localization and non-ergodic behaviour~\cite{Huse2013,
  Tieleman2013} as originally envisaged by Anderson~\cite{Anderson} in
the context of disordered spin systems in solids.
Indeed, we shall show that in the presence of disorder the collective
dynamics of the rotational levels does exhibit such non-ergodic
behaviour, with a transition from BEC to an uncondensed phase which
does not coincide with the equilibrium phase transition.

We investigate a system of polar molecules prepared initially in their
ro-vibrational groundstate and loaded in a single quasi-2D layer
subject to a deep square optical lattice. The molecules are assumed to
be in a Mott phase with one molecule per lattice site, such that
tunneling of molecules between lattice sites can be neglected.  The
relevant degrees of freedom are the rotational excitations of the
molecules.

We consider the rotational excitations introduced by a resonant
microwave (MW) pulse linearly polarized perpendicular to the 2D
plane. This prepares the initial state
\begin{equation}
|\Psi\rangle=\prod_{i=1}^N\left(\cos\frac{\theta}{2} |\downarrow\rangle_i + \sin\frac{\theta}{2} e^{i\phi} |\uparrow\rangle_i\right)
\label{eq:prepare}
\end{equation}
where $|\downarrow\rangle_i$ is the rotational groundstate ($\ell=0$,
$m_\ell=0$) of the molecule at site $i$ and $|\uparrow\rangle_i$ is
its $\ell=1$, $m_\ell=0$ rotational excited state (with $m_\ell$ the
angular momentum projection normal to the 2D plane).  The
angles $\theta$ and $\phi$ are set by the MW pulse (amplitude and
phase). These are uniform over the system ($\theta_i = \theta$,
$\phi_i=\phi$), giving a uniform density $\rho =
\sin^2\frac{\theta}{2}$ of rotationally excited states with complete
phase coherence.

Dipole-dipole interactions allow resonant exchange of rotational
excitations between pairs of molecules, as described by the
quantum spin Hamiltonian~\cite{Lukin2006,ColoradoExp}
\begin{equation}
\mathcal{H}_{\rm{dip}}=\frac{J_0 a^3}{2}\sum_{i\neq j}\frac{1}{|\mathbf{R}_i-\mathbf{R}_j|^3}(S^-_iS^+_j+S^-_jS^+_i),
\label{eq:ham}
\end{equation}
where $S^{\pm}_i$ are the spin raising/lowering operators for the
spin-1/2, $a$ is the lattice constant, and $J_0 = d^2/4\pi\epsilon_0 a^3$ is the coupling strength for nearest neighbour sites in
terms of the dipole matrix element $d$ between $|\uparrow\rangle$ and $|\downarrow\rangle$ (e.g. $J_0/h = 52$ Hz for
KRb in a lattice with $a = 532$ nm~\cite{ColoradoExp}).  Note that, in 2D the resonant
transfer preserves $m_\ell=0$, so states with $m_\ell=\pm 1$ remain
unpopulated~\cite{Note1}.

Since (\ref{eq:prepare}) is not an eigenstate of the Hamiltonian
(\ref{eq:ham}), the state will evolve in time. Our goal is to
determine the long-time behaviour.  The central question is: does the
system retain long-range correlations in local phases $\phi_i$? If so,
then the system will remain resistant to decoherence, leading to a
divergent $T_2$ coherence time as measured in standard spin-echo
experiments. We will show that dipole-dipole interactions can cause the
system to be resistant to decoherence: both for the ideal
system with only dipole interactions (\ref{eq:ham}); and  for
systems with additional disorder [Eqn.~(\ref{eq:site})], representing inhomogeneous
broadening of the molecular rotational transition, provided this
disorder is below a critical value.

{\it BEC in the disorder-free system} --- To discuss (de)coherence, it is convenient to recast the model in terms of
hard-core bosons, with $|\downarrow\rangle$ an empty site and
$|\uparrow\rangle$ an occupied site.  Within this viewpoint, the
prepared state (\ref{eq:prepare}) is a BEC. The question of survival of long-range
coherence becomes: does the system remain a BEC at long times? We
shall quantify the residual coherence by computing the condensate
fraction.

In the boson picture, the Hamiltonian (\ref{eq:ham}) is the hard-core
limit of the Bose-Hubbard model. However, it has the interesting
feature that the intersite hopping scales with distance as
$1/r^3$.  This leads to a dispersion relation for the bosons
$\epsilon_{\mathbf{k}} = J_0 a^3 \sum_i \frac{e^{i\mathbf{k}\cdot
    \mathbf{R}_i}}{|\mathbf{R}_i^3|}$ for which $\delta\epsilon_{\mathbf{k}} \equiv
\epsilon_{\mathbf{k}}-\epsilon_{\mathbf{0}}
\propto |\mathbf{k}|$ at small wavevectors.  This relativistic
dispersion has striking consequences for the thermodynamic phase
diagram of the model.  Specifically, for a 2D system of conventional
massive particles with $\delta\epsilon_{\mathbf{k}} \propto |\mathbf{k}|^2$
there can be no BEC at non-zero temperature. (Thermal fluctuations
diverge, and leave at most power-law order in the condensate
phase~\cite{Berezinskii, Thouless}.)  However, for relativistic
particles, the density of states vanishes sufficiently rapidly close
to $\mathbf{k}=0$ to allow BEC in 2D even at non-zero
temperature.  As we now show, this BEC phase can be achieved by MW
excitation.

To discuss this, we first make the assumption that the time evolution
of the initial state (\ref{eq:prepare}) is such that at long times the
system thermalizes. (We shall discuss in detail below dynamical issues that
can prevent thermalization.) Since time evolution under the
Hamiltonian (\ref{eq:ham}) conserves both energy and number of
rotational excitations, the final (thermal) state has temperature and
chemical potential for the bosons that are set by the initial
state. Note that, because the energy scale $J_0$ in (\ref{eq:ham}) is
positive, the total energy of the prepared state is of the order of
$J_0$ per lattice site and is also positive. That is, it is close to
the {\it maximum} energy state of (\ref{eq:ham}).  Thus, the
corresponding thermalized states will be at {\it negative} absolute
temperature $T$. The condensed ``groundstate'' that we will expand
around is this maximum energy state, which is reached in the limit
$T\to 0$ from {\it negative} temperatures;  we shall  denote this the ``$T =
0^-$ groundstate''.

The thermodynamic properties for small negative
temperatures can be well described by a spin-wave analysis. At a mean excitation density $\rho$, this analysis leads to the mean energy
\begin{eqnarray}
\langle \mathcal{H}_{\rm{dip}} \rangle_T =  N J \rho (1-\rho) + \sum_{\mathbf{k\neq 0}} \left(E_\mathbf{k} n_{\mathbf{k}}(T)+ F_\mathbf{k}\right) 
\label{Bog}
\end{eqnarray}
where $J=\epsilon_{\mathbf{k=0}}\sim 9.0J_0$, $n_{\mathbf{k}}(T)=[\exp(E_\mathbf{k}/k_B T)-1]^{-1}$ is the thermal boson occupation number,  and $E_\mathbf{k}=-\sqrt{(\epsilon_{\mathbf{k}}(1-2\rho)^2-J)(\epsilon_{\mathbf{k}}-J)}$ and $F_\mathbf{k}=\frac{1}{2}[J+E_\mathbf{k}-\epsilon_{\mathbf{k}}(1-2\rho+2\rho^2)]$ are, respectively, the dispersion of elementary excitations and the zero-point fluctuations. 
Similarly, the condensate density is
\begin{eqnarray}
\rho_0 & \equiv & \sum_{i,j}\frac{ \langle S_{i}^{+}S_j^- \rangle}{N^2}
= \rho(1-\rho) - \frac{1}{N} \sum_{\mathbf{k\neq0}} \left (n^Q_{\mathbf{k}}+n^T_{\mathbf{k}}\right )
\end{eqnarray} 
where $n^Q_{\mathbf{k}}=-\frac{F_\mathbf{k}}{E_\mathbf{k}}$ and $n^T_{\mathbf{k}}=(1-2\frac{F_\mathbf{k}}{E_\mathbf{k}})n_{\mathbf{k}}(T)$ are the numbers of particles depleted
from the condensate into the $\mathbf{k}$ mode due to, respectively, quantum and thermal
fluctuations computed within Bogoliubov theory.
 The state
(\ref{eq:prepare}) has an initial condensate fraction $\rho_0 =
\rho_0^{\rm max} \equiv \rho(1-\rho)$ which is the maximum
  possible for hard-core bosons.  

The energy of the prepared state (\ref{eq:prepare}) is a conserved quantity in the
isolated system and is given by
\begin{equation}
 \langle \mathcal{H}_{\rm{dip}} \rangle_{\rm{prepared}} = N J  \rho (1-\rho).
\label{eq:meanenergy}
\end{equation}
Within the mean-field approximation the prepared state would be the
exact $T=0^-$ groundstate. However, quantum fluctuations cause
the exact $T=0^-$ groundstate to differ (slightly) from this
  prepared state, such that the mean energy (\ref{eq:meanenergy})
  corresponds to a (small) non-zero temperature.  Under the
assumption of thermalization, the prepared states will
  thermalize to the BEC phase provided that $\langle
\mathcal{H}_{\rm{dip}} \rangle_{\rm{prepared}}$ is larger than the
mean energy at the transition point $\langle \mathcal{H}_{\rm{dip}}
\rangle_{T_{\rm C}}$ at which the condensate density vanishes, $\rho_0(T_{\rm C}) = 0$.

\begin{figure}
\centering
\includegraphics[width=1.0\columnwidth, angle=0]{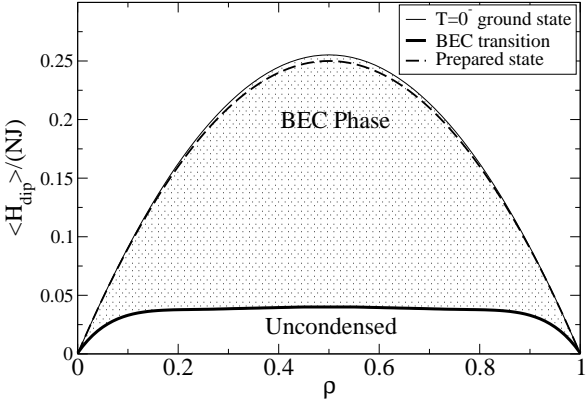}
\caption{Phase diagram of the disorder-free system. The prepared state is always close to the $T=0^-$ groundstate because of small spin-wave corrections and hence thermalizes in the BEC phase for all $\rho$. The BEC transition line is the energy of the system corresponding to the transition temperature.} \label{phases}
\end{figure}
Fig.~\ref{phases} summarizes the results of the above analysis. We
find that the zero-point corrections are small, so the energy of the
prepared system is always very close to that of the $T=0^-$
groundstate. (The smallness of these corrections justifies our
spin-wave approximation.)  Hence, after thermalization, the
system is in the ordered (BEC) phase for all densities $\rho$,
so the microwave pulse 
creates a state with infinitely long-lived coherence.

{\it The disordered system} --- So far we have considered the pure
system without external sources of decoherence. In
experiment it is natural to have sources of
broadening of the rotational transition, from inhomogeneities of
lightshift or random filling of the lattice. In current experiments~\cite{ColoradoExp},
optical lattices are loaded with molecules at filling
fractions much smaller than unity, circa $0.1$. This is the limit
where disorder dominates and leads to decoherence of the prepared state (\ref{eq:prepare}).

Related disordered spin models have been extensively studied in
1D~\cite{Fisher1992, Fisher1994,ColoradoTheory}, where strong quantum
fluctuations preclude any long-range order. Here, we study the effects of disorder in 2D,
where true long-range order can survive. We use a mean-field approach,
motivated by the fact that zero-point corrections are very small for
the clean system, and further supported by beyond-mean-field
treatments described below.
The disorder enters through an onsite coupling
\begin{equation}
\mathcal{H}_{\rm{site}}=-\sum_i (\mu+\xi_i) S^z_i,
\label{eq:site}
\end{equation}  
where $S^z_i=\pm 1/2$ for $|\uparrow\rangle/|\downarrow\rangle$,
$\xi_i$ are Gaussian-distributed quenched random variables of standard
deviation $W$ which sets the scale of inhomogeneous broadening. (For
uncoupled spins, $J_0=0$, the initial state (\ref{eq:prepare}) would
decohere with $T_2 \sim h/W$.)  We have introduced a chemical potential
$\mu$ to fix the density of rotational excitations $\rho$.  Without
loss of generality we take the mean-field to be along the $x$
direction
\begin{equation}
B = \frac{2 J}{N} \sum_i \langle S^x_i\rangle.
\end{equation}
Within mean-field theory the total Hamiltonian $\mathcal{H}_{\rm{site}}+\mathcal{H}_{\rm{dip}}$ is replaced by
\begin{equation}
\mathcal{H}_{\rm{MF}}= \sum _i \mathbf{S}_i \cdot \mathbf{B}_i - \frac{N B^2}{4 J},
\end{equation}
where $\mathbf{B}_i = (B,0,-\mu - \xi_i)$. The spin on each site can
now be treated independently. At equilibrium at temperature
$T$ the mean-field satisfies the self-consistency equation
\begin{equation}
1=J \int p(\xi) \frac{\tanh \left (\frac{1}{2k_B T} \sqrt{B^2 + (\mu + \xi)^2}\right )}{\sqrt{B^2 + (\mu + \xi)^2}} \,d\xi, \label{MF}
\end{equation}
where the summation over sites is equivalent to an ensemble average
over $\xi$ in the thermodynamic limit, with $p(\xi) = (2\pi W^2)^{-1/2}\exp(-\xi^2/2W^2)$.

The solution of Eqn.~(\ref{MF}) shows a phase transition between a BEC
(with $B\neq 0$, giving condensate fraction $\rho_0 = B^2/4J^2$) and uncondensed phases, at a critical temperature that depends on the disorder
strength.
Although mean-field theory shows no phase transition at zero
temperature, in contrast to expectations of more
accurate methods (e.g. \cite{Feigelmann, Ioffe}), this weakness is unimportant for our
purposes, since, as we now show, the transition of interest is in the
regime $J/W > j_{\rm MF}\simeq 0.24$ where mean-field theory is
expected to capture the behaviour accurately~\cite{Feigelmann}.

\begin{figure}
\centering
\includegraphics[width=1.0\columnwidth, angle=0]{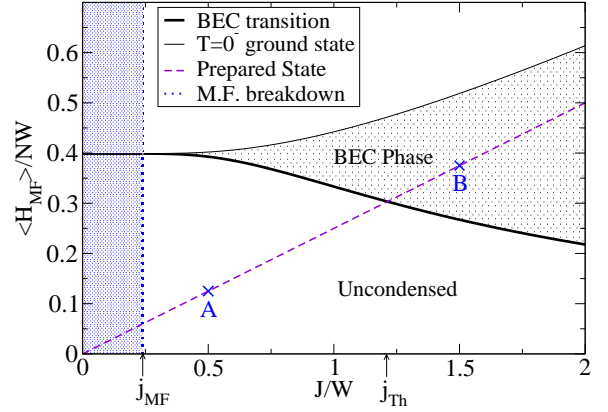}
\caption{Phase diagram of the disordered system for $\rho = \frac{1}{2}$. The BEC transition line is the energy of the system corresponding to the transition temperature. Isolated systems below this energy thermalize in the uncondensed phase. The dashed line is the energy of the prepared state with $\rho = \frac{1}{2}$ which thermalizes in the BEC phase for $J/W > j_{\rm Th}$. The mean-field description breaks down for $J/W < j_{\rm MF} \sim 0.24$. In Fig.~\protect\ref{time} we compare the dynamics at points A and B. } \label{crossover}
\end{figure}

Fig.~\ref{crossover} shows our results for $\rho=\frac{1}{2}$. The
mean energy of the prepared state increases
with $J/W$. For
strong disorder, $J/W < j_{\rm Th}\simeq 1.2$, the mean energy of the
prepared state is such that it is in a regime in which the thermalized
state has no long-range order, so will completely decohere. For weak
disorder, $J/W > j_{\rm Th}\simeq 1.2$, the mean energy of the
prepared state places it in a regime in which the thermalized
state forms a BEC and has long-range coherence. 
Hence, our results show that for sufficient dipolar coupling,
$J/W\gtrsim 1.2$, the thermalized system will remain phase coherent
despite the inhomogeneous broadening ($W\neq 0$).

{\it Dynamical Evolution of the Prepared State} --- Our results have
been based on the assumption that the isolated system will reach
thermal equilibrium at long times, with properties set only by the
initial energy and excitation density $\rho$. We now turn to
investigate the time evolution itself.  We focus on the prepared state
with density $\rho=1/2$, which is representative of other
densities. Within the mean-field model, the dynamics of the
expectation values of the individual spins obey
\begin{figure} \centering
\includegraphics[width=1.0\columnwidth, angle=0]{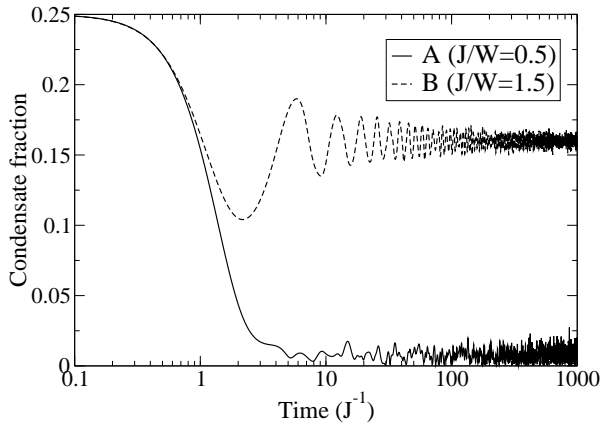}
\caption{Dynamical evolution of the condensate fraction,
computed via $\rho_0 = \frac{ (B_x^2 + B_y^2)}{4 J^2}$ from mean-field theory.   For strong disorder the condensate fraction decays to zero at long times, so the initial state completely decoheres.} \label{time}
\end{figure}
\begin{equation}
\dot{\mathbf{S}_i}= \mathbf{S}_i \times \mathbf{B}_i,
\end{equation}
where $\mathbf{B}_i=(B_x,B_y,\xi_i)$ and $B_x$, $B_y$ are the
mean-fields along the $x$ and $y$ directions respectively.  We
simulate this dynamics numerically for systems of up to $10^5$ spins,
updating the mean-$B_x$ and $B_y$ fields at each time step.

\begin{figure}
\centering
\includegraphics[width=1.0\columnwidth, angle=0]{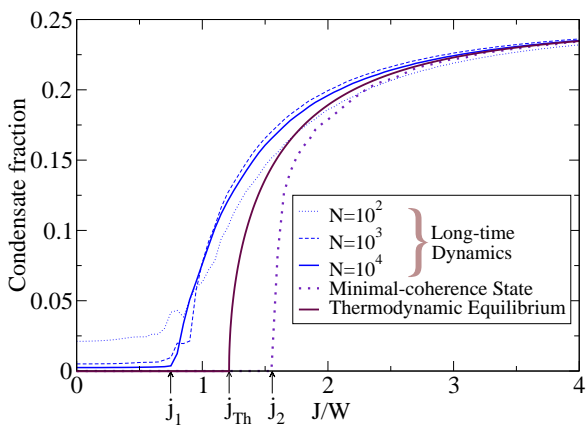}
\caption{Comparison of the condensate fraction for the thermalized
  state with that found from the long-time dynamics of mean-field
  theory (for increasing systems sizes $N$) for a system prepared with
  a coherent microwave pulse.  The prepared state retains coherence in
  a regime $j_{1} < J/W < j_{\rm Th}$ for which the thermalized state
  would be completely incoherent, showing non-ergodic dynamics in the
  disordered system. The dot-dashed line is the long-time behaviour
  for a system prepared in a state of minimal coherence, described in
  the text, with the same mean energy. }
\label{comparison}

\end{figure}

Fig.~\ref{time} shows our results for two disorder strengths. Far inside the region
where the thermalized state is uncondensed (A), the long time dynamics
is to a state with vanishing condensate fraction: i.e. the prepared
state decoheres.  Far inside the region where the
thermalized state would be a BEC (B), the long-time dynamics shows a
reduced, but non-zero residual condensate fraction: i.e. the system
retains coherence to long times. 

Thus we find that the dynamical evolution exhibits a transition
  between condensed (BEC) and uncondensed phases as a function of $J/W$, as we
  had also found based on the thermalization {\it ansatz}. However,
the locations of these transitions differ.
Fig.~\ref{comparison} compares the long-time results of our dynamical simulation
to the predictions based on thermodynamic equilibrium~\cite{Note2}.
Thermodynamic and dynamical results converge for large and small
$J/W$. However, they differ in the intermediate regime. For $j_{\rm 1}
< J/W < j_{\rm Th}$ the long-time dynamics of the system show
long-range order, even though the thermalized system is disordered.
This is a signature of many-body localization, and localization
protected quantum order~\cite{Huse2013}.

That the system does not reach thermal equilibrium for intermediate
$J/W$ shows that the dynamics are non-ergodic, and sensitive to
initial conditions. We have explored this further by simulating the
evolution of a range of systems with different initial states of the same
total energy: from the prepared state with maximal coherence
(\ref{eq:prepare}); to a state of minimal coherence in which $S^z_i$
has the same sign as $\xi_i$ and we choose the maximum $|S^z_i|$
allowed by energy conservation. The dynamics of this state of minimal
coherence also shows non-ergodic behaviour.  In the intermediate
range $j_{1} < J/W < j_{2}$ the system is locked in the
initial state, be it coherent or incoherent, depending on the initial
state (maximum or minimum coherence) and irrespective of the form of
the thermalized state.  Note that the mean-field theory we study is a
{\it classical} non-linear dynamical system.  Similar non-ergodic
behaviour, and lack of thermalization, has also been found in recent
work by Tieleman {\it et al.}~\cite{Tieleman2013}, who have shown that
in the closely related mean-field theory of the Bose-Hubbard model,
disorder breaks the connection between chaoticity and ergodicity even
when classical localization cannot occur. We interpret the non-ergodic
nature of the mean-field spin system as indicative of the many-body
localization in the underlying quantum spin system.

In summary, we have shown that due to dipole-dipole
  interactions the rotational excitations of polar molecules in
2D optical lattices can form a BEC even at non-zero
temperature.  This collective many-body phase gives rise to
  long-lived coherence of the rotational excitations, even in the
  presence of disorder that would otherwise cause inhomogeneous
  broadening. Our results show that experimental studies of the
  (de)coherence of the rotational excitations prepared by a microwave
  pulse, can be used to explore regimes of many-body localization,
  where non-ergodic behaviour allows long-range phase coherence even
  when the thermalized phase would be incoherent.

\acknowledgments{We gratefully acknowledge financial support from
  EPSRC, Grant No. EP/J017639/1.}



\end{document}